\catcode`\@=11

\newif\if@fewtab\@fewtabtrue

{\count255=\time\divide\count255 by 60
\xdef\hourmin{\number\count255}
\multiply\count255 by-60\advance\count255 by\time
\xdef\hourmin{\hourmin:\ifnum\count255<10 0\fi\the\count255}}
\def\ps@draft{\let\@mkboth\@gobbletwo
    \def\@oddhead{}
    \def\@oddfoot {\hbox to 7 cm{\tiny \versionno
       \hfil}\hskip -7cm\hfil\rm\thepage \hfil}
    \def\@evenhead{}\let\@evenfoot\@oddfoot}

\def\draftcite#1{\ifnum\draftcontrol=1#1\else{}\fi}

\def\@lbibitem[#1]#2{\item{}\hskip -3cm \hbox to 2cm
{\hfil$\scriptstyle\draftcite{#2}$}\hskip
1cm[\@biblabel{#1}]\if@filesw
     {\def\protect##1{\string ##1\space}\immediate
      \write\@auxout{\string\bibcite{#2}{#1}}}\fi\ignorespaces}

\def\@bibitem#1{\item\hskip -3cm \hbox to 2cm
{\hfil {\footnotesize\draftcite{#1}}}\hskip 1cm
\if@filesw \immediate\write\@auxout
       {\string\bibcite{#1}{\the\value{\@listctr}}}\fi\ignorespaces}

\def\yes{yes }
 \read-1 to\yesno
\ifx\yesno\yes 
\font\tendl=msym10  scaled \magstep1
\font\sevendl=msym7 scaled \magstep1
\font\fivedl=msym5 scaled \magstep1
\newfam\dlfam \def\dl{\fam\dlfam\tendl}
\textfont\dlfam=\tendl \scriptfont\dlfam=\sevendl
\scriptscriptfont\dlfam=\fivedl
\else \let\dl=\bf
\fi

\global\def\draftcontrol{0}
\catcode`\@=12

\def\ifundefined#1{\expandafter\ifx\csname#1\endcsname\relax}
\makeatletter \ifundefined{new@mathgroup} {} \else
    \new@mathgroup\sffam
    \new@internalmathalphabet\mathsf\sffam{cmss}{m}{n}
    \def\psf{\fontfamily\sfdefault \fontseries\default@series
        \fontshape\default@shape\selectfont\mathsf}
\fi

\def\citen#1{\if@filesw \immediate\write \@auxout {\string\citation{#1}}\fi%
\@tempcntb\m@ne \let\@h@ld\relax \def\@citea{}%
\@for \@citeb:=#1\do {\@ifundefined {b@\@citeb}%
    {\@h@ld\@citea\@tempcntb\m@ne{\bf ?}%
    \@warning {Citation `\@citeb ' on page \thepage \space undefined}}%
    {\@tempcnta\@tempcntb \advance\@tempcnta\@ne
    \setbox\z@\hbox\bgroup\ifcat0\csname b@\@citeb \endcsname \relax
    \egroup \@tempcntb\number\csname b@\@citeb \endcsname \relax
    \else \egroup \@tempcntb\m@ne \fi \ifnum\@tempcnta=\@tempcntb
    \ifx\@h@ld\relax \edef \@h@ld{\@citea\csname b@\@citeb\endcsname}%
    \else \edef\@h@ld{\hbox{--}\penalty\@highpenalty
    \csname b@\@citeb\endcsname}\fi
    \else \@h@ld\@citea\csname b@\@citeb \endcsname \let\@h@ld\relax \fi}%
\def\@citea{,\penalty\@highpenalty\hskip.13em plus.13em minus.13em}}\@h@ld}
\def\@citex[#1]#2{\@cite{\citen{#2}}{#1}}%
\def\@cite#1#2{\leavevmode\unskip\ifnum\lastpenalty=\z@\penalty\@highpenalty\fi%
  \ [{\multiply\@highpenalty 3 #1%
  \if@tempswa,\penalty\@highpenalty\ #2\fi}]}   %
\makeatother 

\def\alg     {algebra}

\newcommand{\authoretc}[5]{\centerline{\sc #1}\vskip2 mm
                   \centerline{#2}\vskip.5mm \centerline{#3}\vskip.5mm
                   \centerline{#4}\vskip.5mm \centerline{#5}}
\def\be            {\begin{equation}}

\def\cft           {conformal field theory}
\def\cfts          {conformal field theories}
\def\class   {classification}

\def\con     {conformal\ }
\def\Con     {Conformal\ }

\def\ee            {\end{equation}}

\let\emb=\hookrightarrow
\newcommand{\erf}[1]{(\ref{#1})}
\def\findim        {finite-dimensional}
\newcommand{\fline}[1]{\vskip 4mm\noindent ------------------\\[1 mm]}
\def\furu    {fusion rule}
\def\futnote#1     {\footnote{~#1}\ }
\def\g             {{\sf g}}
\def\gagr          {Galois group}

\def\gv            {g_{}^\vee}
\def\h             {{\sf h}}

\newcommand{\hsp}[1] {\mbox{\hspace{#1 em}}}

\def\hy            {$\mbox{-\hspace{-.66 mm}-}$}
\def\id            {{\sl id}}
\def\ii            {{\rm i}}

\def\infdim        {infinite-dimensional}
\def\Infdim        {Infinite-dimensional}
\def\inv           {invariance}

\def\jf            {J.\ Fuchs}
\def\km            {Kac\hy Moo\-dy }
\def\kma           {Kac\hy Moo\-dy algebra}

\long\def\labl#1   {\ifnum\draftcontrol=0 \label{#1}\ee \else \label{#1}\ee
                   \mbox{ }\\[-12 mm]\query{#1}\\[5 mm] \fi}
\def\lie           {Lie algebra}

\def\Modinv  {Modular invarian}
\def\modinv  {modular invarian}

\def\mydollar      {$^\pounds$} 
\def\one           {\mbox{\small $1\!\!$}1}

\def\parfu   {partition function}
\def\qdim          {quantum dimension}

\long\def\query#1{\hskip 0pt{\vadjust{\everypar={}\small\vtop to 0pt{\hbox{}%
     \vskip -13pt\rlap{\hbox to 49.9pc{\hfil{\vtop{\hsize=8pc\tolerance=6000%
     \hfuzz=.5pc\rightskip=0pt plus 3em\noindent#1}}}}\vss}}}}%

\def\rationals     {{\dl Q}}

\def\rep           {representation}

\newcommand{\sect}[1] {\section{#1}\setcounter{equation}{0}}
\def\sign          {\mbox{sign\,}}

\def\smat          {$S$-matrix}
\def\sym     {symmetry}

\newcommand{\version}[1] {\ifnum\draftcontrol=0 {} \else
                   \typeout{}\typeout{#1}\typeout{}\vskip3mm
                   \centerline{{\tt DRAFT -- #1 -- \today}} \vskip3mm \fi}

\def\wzwm    {WZW model}
\def\wzwt          {WZW theory}
\def\wzwts         {WZW theories}
\def\zet           {{\dl Z}}

\def\zetplus       {\mbox{${\zet}_{>0}$}}

\def\zett          {\mbox{\small {\dl Z}}}

   \newcommand{\wb}{\,\linebreak[0]} 
   \def\wB                {$\,$\wb}
   \newcommand{\Bi}[1]    {\bibitem{#1}}

   \newcommand{\vypf}[5]  {\ {#1} [FS{#2}] ({#3}) {#4}}
   \newcommand{\J}[5]     {\ {\sl #5}, {#1} {#2} ({#3}) {#4}}
   \newcommand{\Prep}[2]  {{\sl #2}, preprint {#1}}

   \def\adma  {Adv.\wb Math.}

   \def\comp  {Com\-mun.\wb Math.\wb Phys.}

   \def\ijmp  {Int.\wb J.\wb Mod.\wb Phys.\ A}

   \def\npbF  {Nucl.\wb Phys.\ B\vypf}

   \def\nupb  {Nucl.\wb Phys.\ B}

   \def\phlb  {Phys.\wb Lett.\ B}

   \def\phrd  {Phys.\wb Rev.\ D}

   \def\pnas  {Proc.\wb Natl.\wb Acad.\wb Sci.\wb USA}

   \def\rims  {Publ.\wB RIMS}

\documentstyle[11pt]{article}

\ifx\yesno\yes 
\else  \fi
\def\BB            {{\cal B}}
\def\calP          {{\cal Z}_{{\rm so}(d)}}
\def\caLP          {\hat{\cal Z}_{{\rm so}(d)}}
\def\calX          {\chi}
\def\calY          {\chi}
\def\calYi         {\chi^{}_i}

\def\Chi           {{\cal X}}
\def\chic          {\Chi_{\rm c}}
\def\chio          {\Chi_{\rm o}}
\def\chis          {\Chi_{\rm s}}
\def\chiv          {\Chi_{\rm v}}
\def\cft           {conformal field theory}
\def\cfts          {conformal field theories}
\def\Cr            {\\[1.2mm]}

\def\D             {D}
\newcommand{\deltal}[2] {\delta^{[\el]}_{#1,#2}}
\newcommand{\deltalr}[2] {\delta^{[\el\rho]}_{#1,#2}}
\newcommand{\deltar}[2] {\delta^{[\rho]}_{#1,#2}}
\let\dstyle=\displaystyle
\def\dprod         {\displaystyle\prod}

\def\EE            {{\rm E}}
\def\el            {\ell}
\def\Ell           {p}
\def\elm           {\ell^{-1}}
\def\elr           {\ell^{\,r}}
\def\elre          {\ell^{\,r-1/2}}
\def\epsl          {\epsilon_{\el}}

\def\epsz          {\epsilon}
\newcommand{\fI}[1]{\varphi_{#1}^{}}
\newcommand{\Fi}[1]{\phi_{#1}^{}}
\def\findim        {finite-dimen\-sional}
\def\gagr          {Galois group}

\def\gb            {\mbox{$\bar{\sf g}$}}
\def\gv            {{g^\vee}}
\def\h             {{\sf h}}
\def\half          {\mbox{$\frac12\,$}}
\def\ii            {{\rm i}}
\def\Kacg          {\g}
\def\kg            {h}
\def\kv            {{k^{\vee}}}

\def\llb           {\mbox{\large(}}

\def\lkg           {\el\kg}
\def\lrb           {\mbox{\large)}}

\def\lv            {\mbox{$L^\vee$}}
\def\mo            {\mbox{${\rm m}_{\oo}$}}

\def\moi           {\mbox{${\rm m}_{\oo}^{\,i}$}}
\def\mod           {\mbox{mod\,}}

\def\mOO           {\mbox{${\rm m}_{0}^{\,0}$}}
\def\mor           {\mbox{${\rm m}_{\oo}^{\,\rho}$}}
\def\mov           {\mbox{$\vec{\rm m}_{\oo}$}}
\def\mpe           {\mbox{${\rm m}_{\Ell}$}}
\def\mpi           {\mbox{${\rm m}_{\Ell}^{\,i}$}}
\def\mpI           {\mbox{${\rm m}_{{\Ell}'}^{\,i}$}}
\def\mpj           {\mbox{${\rm m}_{\Ell}^{\,j}$}}
\def\mpJ           {\mbox{${\rm m}_{{\Ell}'}^{\,j}$}}

\def\mv            {\mbox{${\rm m}_{\rm v}$}}
\def\mvi           {\mbox{${\rm m}_{\rm v}^{\,i}$}}
\def\mvr           {\mbox{${\rm m}_{\vv}^{\,\rho}$}}
\def\mvv           {\mbox{$\vec{\rm m}_{\vv}$}}
\def\nn            {{d/2}}
\def\nN            {d}
\def\np            {\overline{\cal N}}

\def\oo            {{\rm o}}
\def\OO            {0}
\def\P             {P}
\def\Pr            {Q}
\def\pk            {{P_\kg}}
\def\pkg           {\mbox{\large$\frac{2\pi\ii}{\kg}$}}
\def\pkl           {\mbox{\large$\frac{2\pi\ii}{\el}$}}
\def\plh           {{P_{\el\kg}}}
\def\plkg          {\mbox{\large$\frac{2\pi\ii}{\lkg}$}}
\def\qdim          {quantum dimension}
\def\qg            {quasi-Galois}
\def\rcft          {rational \cft}
\def\rcfts         {rational \cfts}
\def\rh            {\mbox{${\cal R}_{\kg}$}}
\def\rhO           {\mbox{$L_{\rho}$}}
\def\rlh           {\mbox{${\cal R}_{\lkg}$}}
\def\semitimes     {\begin{picture}(9.4,8)\put(1.57,0.32){\line(0,1)
                     {5.4}}\put(0,0){$\times$} \end{picture} }
\def\Sig           {\Sigma^{-1}}
\def\Siglr         {\Sigma^{-1}(\el\rho)}
\def\Siglrm        {\tilde\Sigma^{-1}(\el\rho)}
\def\SiglRm        {\tilde\Sigma^{-1}(\rho)}

\def\SIglrmm       {{\mu\in\Sigma^{-1}(\rho)\atop\epsz(\mu)=-1}}

\def\SiglrMM
{{\mu,\mu'\in\Sigma^{-1}(\rho)\atop\epsz(\mu)=\epsz(\mu')=-1}}

\def\SIglrmp       {{\mu\in\Sigma^{-1}(\rho)\atop\epsz(\mu)=1}}

\def\SiglrMP
{{\mu,\mu'\in\Sigma^{-1}(\rho)\atop\epsz(\mu)=\epsz(\mu')=1}}
\def\sigmA         {\dot\sigma}

\def\sigmAl        {\dot\sigma_{(\el)}^{}}

\def\sigmAm        {\dot\sigma^{-1}}
\def\sign          {\mbox{sign}}

\def\signw         {\mbox{sign}(w)}
\def\signwa        {\mbox{sign}(w_A^{})}
\def\signwc        {\mbox{sign}(w_C^{})}
\def\Sigzr         {\Sigma^{-1}(\rho)}
\def\sm            {\mbox{\rm s}}
\def\SM            {\mbox{\rm S}}
\newcommand{\sM}[2] {\mbox{{\rm s}$_{#1,#2}$}}
\newcommand{\Sm}[2] {\mbox{$S_{#1,#2}$}}
\newcommand{\Smt}[2] {\mbox{{\rm S}$_{#1,#2}$}}
\def\smat          {$S$-matrix}

\def\sod           {\mbox{so($d$)}}
\def\SOD           {\mbox{$\widehat{\rm so}(d)$}}
\newcommand{\Ss}[2] {\mbox{${\cal S}^{(\el)}_{#1,#2}$}}

\newcommand{\sump}[1] {\sum_{#1\in\pk}}
\newcommand{\sumP}[1] {\sum_{#1\in\plh}}

\def\sumpc         {\sum_{c\in P_k}}
\newcommand{\sumt}[2] {\sum_{#1\in\T(#2)}}
\def\sumts         {\sum_{s\in\T}}
\def\sumtT         {\sum_{t\in\T}}
\def\sumtt         {\sum_{s\in\TT}}
\def\sumww         {\sum_{w\in W}}
\def\T             {{\cal T}}

\def\TT            {\tilde{\cal T}}
\def\vecv          {\vec{\rm e}}
\def\vv            {{\rm v}}
\def\wh            {\hat w}

\def\wha           {\wh_a}
\def\wl            {\hat W_{\el}}
\def\wzwt          {WZW theory}
\def\wzwts         {WZW theories}
\def\zce           {Z_{\rm c.e.}}
\def\zCe           {Z'_{\rm c.e.}}
\def\zcE           {({Z_{\rm c.e.}})}
\def\zCE           {({Z'_{\rm c.e.}})}
\def\zcec          {{\cal Z}_{\rm c.e.}}
\def\zh            {\hat Z}
\def\zt            {\tilde z}
\def\ztt           {\tilde{\!\tilde z}}
\def\Zt            {\tilde Z}
\def\Ztt           {\tilde{\!\tilde Z}}

\begin{document} 

\vspace*{1cm}
\begin{flushright}  {~} \\[-26 mm] {\sf hep-th/9412009}\\ 
{\sf NIKHEF-H/94-37} \\[1 mm]{\sf December 1994} \end{flushright}

\vskip 15mm
\begin{center}
{\Large{\bf QUASI-GALOIS SYMMETRIES}} \vskip 4mm
{\Large{\bf OF THE MODULAR $S$-MATRIX}} \end{center}
\vskip 18mm
 \authoretc{J\"urgen Fuchs, \mydollar \ \, Bert Schellekens, \
 Christoph Schweigert}
 {~} {NIKHEF-H\,/\,FOM} {Kruislaan 409} {NL -- 1098 SJ~~Amsterdam}{~}
\vskip 13mm

\begin{quote} {\bf Abstract.} \\
The recently introduced Galois symmetries of \rcft\ are generalized, for the
case of \wzwts, to `quasi-Galois symmetries'. These symmetries
can be used to derive a
large number of equalities and sum rules for entries of the modular matrix $S$,
including some that previously had been observed empirically. In addition,
quasi-Galois symmetries allow to construct modular invariants and to relate
$S$-matrices as well as modular invariants at different levels.
They also lead us to an extremely plausible conjecture for the branching
rules of the conformal embeddings $\g\emb\widehat{\rm so}({\rm dim}\,\g)$.
\end{quote}

\vfill
{}\fline{} {\small \mydollar~~Heisenberg fellow} \newpage

\sect{Introduction}

In the study of rational conformal field theories, modular transformations
play an essential role. They turn the set of the characters of all primary
fields into a unitary module of $SL(2,\zett)$, the twofold covering of the
modular group of the torus. Via the Verlinde formula, they are also closely
related to the fusion rules.

In all cases where the modular matrix $S$ is explicitly known, one observes
that it contains surprisingly few different numbers, and that among the
distinct numbers there are
linear relations. While it has been known for a long time that simple currents
lead to relations between individual \smat\ elements \cite{scya,scya6,intr},
many other relations, in particular sum rules, have remained so far somewhat
mysterious. Recently it has become clear that Galois symmetries
\cite{dego,coga} are an independent
source for relations between individual elements of $S$ \cite{fgss,fusS}.
Both simple current and Galois symmetries exist for arbitrary \rcfts,
independent of the structure of the chiral algebra.

In this paper we will show
that in the special case of \wzwts, Galois symmetries can be generalized to
what
we will call {\em quasi-Galois symmetries}. A crucial ingredient of our
construction (which is not available for other \cfts\ than \wzwts) is the
Kac\hy Peterson formula for the \smat.
These new symmetries turn out to be rather powerful and allow to derive three
new types of relations between the entries of $S$: first, a sum rule which
relates
signed sums of \smat\ elements, see \erf1; second, the equality, modulo signs,
of certain specific \smat\ elements, see \erf{Slac}; third,
a new systematic reason for \smat\ elements to vanish, see the remarks
after \erf{01}.

Just as in the case of Galois symmetries, the relations we find can be
employed to construct
elements of the commutant of $S$, and therefore to generate modular
invariants. Moreover, they can be used to obtain relations between
invariants at different values of the level, i.e.\ between different \wzwts.
Finally, we present arguments that our results allow to determine
the branching rules of certain conformal embeddings.

The rest of the paper is organized as follows. In section 2 we recall the
basic facts about Galois symmetries of \rcfts, and of \wzwts\ in
particular, and show how in the WZW case they can be generalized to \qg\
symmetries. Also, as a first application, we describe how these symmetries
force certain \smat\ elements to vanish. Section 3 contains the proof of
the sum rule \erf1 for the entries of $S$, and in section 4 this sum rule
is used to construct integral-valued matrices that commute with the \smat.
In section 5 we obtain another symmetry, \erf{Slac}, of $S$ as well as
relations (see \erf i, \erf{ii}) between the $S$-matrices for \wzwts\
at different heights $h_1,\;h_2$, where $h_1$ is a multiple of $h_2$. Again,
these
results lead to a prescription for constructing \smat\ invariants, now
both at the smaller and at the larger height (see \erf{zt} and \erf{Zt},
respectively). Finally, in section 6 we consider a special case of the
latter invariants, which leads us to a conjecture for
the branching rules of certain conformal embeddings, and we show that
this conjecture passes various consistency checks.

\sect{Quasi-Galois scalings}

When analyzing the mathematical structure of a \wzwt, we are dealing with
integrable highest weight \rep s of an untwisted affine \lie\ g at a
fixed integral level $\kv$. As the level
is fixed, the g-weights are already fully determined by their horizontal part,
i.e.\ by the weight with respect to the horizontal subalgebra $\gb$ of \g.
In the following it will be convenient to shift all weights according to
  $    a\;\hat=\;\lambda_a+\rho  $
by the Weyl vector $\rho$. Note that if the non-shifted weight $\lambda_a$ is
at level $\kv$, the shifted weight $a$ is at level $\kg$, where
  \be  \kg := \kv+\gv  \labl{kg}
with $\gv$ the dual Coxeter number of $\gb$; we will call $\kg$ the {\em
height\/}
of the weight $a$. The set of (shifted) integrable weights of the affine Lie
algebra \g\ at height $\kg$ is
  \be  \pk:=\{ a \in L^{\rm w} \mid 0<a^i\le \kv+1\ {\rm for}\ i=0,1,...\,,r \}
  \,. \labl{pk}
Here $L^{\rm w}$ denotes the weight lattice, i.e.\ the \zett-span of the
fundamental weights. 
In other words, the weights \erf{pk} are precisely the integral weights in
the interior of the dominant affine Weyl chamber at level $\kv+\gv$.

An important tool for studying the modular properties of \wzwts\
is the Kac\hy Peterson formula \cite{kape3}
  \be  \Sm ab={\cal N} \sumww \signw \, \exp[-\pkg\,(w(a),b)]  \labl{kp}
for the modular matrix $S$. Here the summation is over the Weyl group $W$ of
the \findim\ horizontal subalgebra $\gb$ of \g.
Some immediate consequences of this formula are the following. First,
the fact that according to \erf{kp} $\Sm ab$ depends on $a$ and
$b$ only via the inner products $(w(a),b)$ and the identity
$(w(\el a),b)=\el\,(w(a),b)=(w(a),\el b)$ imply that
  \be  \Sm{\el a}b = \Sm a{\el b}  \,;  \labl{slab}
and second, for any element $\hat w$ of the affine Weyl group $\hat W$
(i.e.\ the horizontal projection of
the Weyl group of the affine algebra \g), one has
  \be  \Sm{\hat w(a)}b = \sign(\hat w)\,\Sm ab  \,.  \labl{swab}
This implies in particular that $\Sm ab=0$ whenever $a$ or $b$ lies on
the boundary of an affine Weyl chamber. Note that in \erf{slab} and \erf{swab}
it is implicit that the quantity $\Sm ab$ given
by \erf{kp} can be considered also for weights which are not integrable.
This is possible because we are free to take the formula \erf{kp}
(which for integrable weights yields the entries of the actual \smat,
i.e.\ of the matrix which realizes the modular transformation $\tau\mapsto-1/
\tau$ on the characters) for arbitrary weights $a,\;b$ as the definition
of $\Sm ab$. Analogously, these weights need not even be integral, and hence
\erf{slab} is valid for arbitrary numbers $\el$, not just for integers.

To apply Galois theory to \cft, one
considers the number field that is obtained as the extension of the rationals
\rationals\ by all \smat\ elements. One can show \cite{coga} that this
extension is a Galois extension and that its \gagr\
is abelian, implying that the number field
is contained in some cyclotomic field $\rationals(\zeta_n)$.
The \gagr\ of the extension $\rationals(\zeta_n)/\rationals$
is isomorphic to $\zett_n^*$, the multiplicative group of all
elements of $\zett_n$ that are coprime with $n$. The Galois automorphism
corresponding to an element $\el\in\zett_n^*$ acts as $\zeta_n\mapsto
(\zeta_n)^\el$.

In the special case of the \wzwt\ based on the untwisted affine Lie algebra \g\
at
height $\kg$, the relevant root of unity is given by $\zeta_{M\kg}$, with $M$
the smallest positive integer for which the $M$-fold of all entries of the
metric on the weight space of \gb\ is integral.
 \futnote{Actually the cyclotomic field $\rationals(\zeta_{M\kg})$
does not yet always contain the normalization
$\cal N$ appearing in \erf{kp}; rather, sometimes
a slightly larger cyclotomic field must be used \cite{coga}.
However, the permutation $\sigmA$ can already be determined from the
generalized
quantum dimensions, which do not depend on $\cal N$.
Accordingly, the correct Galois treatment of $\cal N$ just
amounts to an overall sign factor which is irrelevant for our purposes.}
 A Galois transformation labeled by $\el\in\zett_{M\kg}^*$ then induces
the permutation $\Lambda\mapsto\wh(\el(\Lambda+\rho))-\rho$ of the highest
weights carried by the primary WZW fields, or equivalently, the permutation
  \be  \sigmA\equiv\sigmAl:\quad a \mapsto \sigmA a:=\wha(\el a) \,
  \labl{wla}
of shifted highest weights.
Here $\wha$ is an element of the affine Weyl group at level $\kg$, i.e.
  \be  \wha(b)= w_a(b) + \kg\,t_a \,,  \labl{wha}
where $w_a$ is some element of the finite Weyl group $W$ and $t_a$ some weight
which belongs to the coroot lattice \lv\ of \gb. They are defined by the
condition that $\wha(\el a)\in\pk$, which determines $w_a$ and $t_a$ uniquely.
Substituting \erf{wla} into the formula for WZW conformal dimensions one
easily obtains a condition for $T$-invariance, namely
$\el^2=1$ mod $2M\kg$ (or mod $M\kg$ if all integers $M(a,a)$ are even).
 \futnote{For more details, see in particular appendix A of \cite{fusS}.}

The key idea in the present paper is to allow in the transformation \erf{wla}
for arbitrary integers $\el$ rather than only elements of $\zett_{M\kg}^*$.
As we will show, these generalized transformations lead to interesting new
information. Note that if $\el\not\in\zett_{M\kg}^*$, then in order for the
map \erf{wla} of the integrable weights to be still well-defined, we must
slightly extend the prescription for the Weyl group element $\hat w_a$.
Namely, $\hat w_a$ is now determined by the condition that either
$\el a$ lies on the boundary of some affine Weyl chamber (in which case
$\hat w_a$ can simply be taken to be the identity), or else that
$\wha(\el a)\in\pk$. In the latter case, $\hat w_a$ is the unique element of
$\hat W$ with this property, and we write
  \be  \sign(\wha) =\sign(w_a) =: \epsl(a)  \,, \labl{epsl}
while in the former case we put $\epsl(a)=0$.
While the map \erf{wla} is thus still well-defined for
$\el\not\in\zett_{M\kg}^*$,
it can no longer be induced by a mapping
$\zeta_{M\kg}\mapsto(\zeta_{M\kg})^\el$
of the number field, and hence in particular it does no longer correspond to
a Galois transformation. Nevertheless
the similarity with Galois transformations is still so close that we call the
map
$a\mapsto\el a $, with $\el$ not coprime with $M\kg$, a {\em \qg\ scaling\/}
and the associated map $\sigmA$ \erf{wla} a {\em \qg\ transformation}.

Note that it is not true that an arbitrary integral weight $b$ can be
mapped into $\pk$ by an appropriate affine Weyl transformation. However,
if $b$ is of the special form
$b=\el a$ with $a\in\pk$ and $\el$ coprime with $L\kg$, this is indeed possible
\cite{fusS}; here $L$ denotes the `lacedness' of $\gb$, i.e.
$L=2$ for $\gb$ of type $B$ or $C$ or $F_4$, $L=3$ for $\gb=G_2$, and $L=1$
else.
The condition that $\el$ is coprime with $L\kg$ is in particular fulfilled
whenever
the scaling corresponds to an element of the \gagr, and hence in the case of
genuine Galois transformations a suitable unique $\wha\in\hat W$ exists for any
$a\in\pk$, implying that the map $\sigmA$ is indeed a permutation
of the weights in $\pk$.
In contrast, for a \qg\ scaling there will in general exist some $a\in\pk$
for which $\el a$ lies on the boundary of an affine Weyl chamber, so that
$\sigmA$ is not even an endomorphism of the set of integrable weights.
However, in terms of WZW primary fields the latter situation corresponds to
mapping the primary field with highest weight $a$ to zero, so that $\sigmA$
can still be interpreted as a linear map on
the fusion ring that is spanned by the primary fields. Moreover, this can also
be
translated back to the language of weights by adding to the set $\pk$ a single
element $\BB$ which stands for the union of all boundaries of affine Weyl
chambers. In this setting, the map \erf{wla} supplemented by $\sigmA(\BB)=\BB$
is an endomorphism of the set $\pk\cup\{\BB\}$, though it is not any more
a permutation.

Consider now an arbitrary scaling
  $    a \mapsto \el a \,, \ \el\in\zett\setminus\{0\}\,,  $
with associated (quasi-)\,Galois transformation given by \erf{wla}.
As follows immediately by applying the identities \erf{slab} and \erf{swab}
to $\Sm{\sigmA a}b$, we then have the identity
  \be  \epsl(a)\, \Sm{\sigmA a}b = \epsl(b)\, \Sm a{\sigmA b}  \,.\labl{01}
For genuine Galois scalings, this result was already obtained in \cite{coga}.
In the \qg\ case, the two sides of \erf{01} are not necessarily non-vanishing,
and this provides us with an explanation for the vanishing of certain \smat\
elements. Namely, if for the \qg\ scaling $\el$ the weights
$b$ and $c:=\sigmA a$ are contained in $\pk$, but $\sigmA b$ is not (i.e. $\el
b$
lies on the boundary of an affine Weyl chamber), then \erf{01} tells us that
$\Sm cb=0$. (Another
systematic reason for \smat\ elements to be zero is provided by simple current
symmetries: $\Sm ab=0$ if $a$ is a fixed point of the simple current $J$
and $b$ has non-vanishing monodromy charge with respect to $J$.)

\sect{A sum rule for \smat\ elements}

In this section we will prove that
the following sum rule for the \smat\ elements is valid for all $a,b\in\pk$:
  \be  \sumpc \epsl(c)\, \delta_{a,\sigmA(c)} \Sm cb =
  \sumpc \epsl(c)\, \delta_{b,\sigmA(c)} \Sm ac \,, \labl1
with $\sigmA$ as defined in \erf{wla} and $\epsl$ as in \erf{epsl}.
In the following section we will see that this sum rule can be employed to
construct elements of the commutant of $S$. Generically the sums appearing
in \erf1 contain more than one non-vanishing term; to our knowledge
it is the first time that a relation of this type between \smat\ elements has
been established in a general framework.

By introducing the pre-images of a \qg\ transformation,
  \be  \Sig(a):= \{ c\in\pk \mid \sigmA(c)=a \}  \ee
for any $a\in\pk$, we may rewrite the equality \erf1 in the more suggestive
manner
  \be  \sum_{c\in\Sig(a)} \epsl(c)\, \Sm cb
  = \sum_{c\in\Sig(b)} \epsl(c)\, \Sm ac \,. \labl{1a}
If the map \erf{wla} is invertible, then \erf{1a} reduces to the relation
  \be  \epsl(\sigmAm a)\, \Sm{\sigmAm a}b = \epsl(\sigmAm b)\,
  \Sm a{\sigmAm b}   \,, \labl{es}
which is equivalent to the identity \erf{01} applied to the map $\sigmAm$.

The rest of this section will be devoted to proving the sum rule \erf1.
The proof uses only basic properties of the \smat\ and of
Weyl transformations. However, it is somewhat technical, and since the
manipulations performed in the proof are not essential for most of what
follows, the reader might prefer to skip the rest of this section in a first
reading.
\medskip

To present the proof of \erf1, we need still a bit more notation. First, we
introduce the finite index subgroup
  \be  \wl:= W \semitimes\, \el\kg\lv  \ee
of the affine Weyl group $\hat W=W\semitimes\kg\lv\equiv\hat W_1$.
Factoring out $\wl$ from $\hat W$, one has
  \be  \frac {\hat W} {\wl} = \frac {\kg\lv} {\el\kg\lv} \,, \ee
and a set of representatives of this quotient group is given by
  \be  \T \equiv \T_{(\el)} :=
  \{ \kg t \mid t\in\lv \,;\ 0\leq t_i<\el\ \,\forall\, i=1,2,...\,,r \} \,.
\ee
Further, for any $a\in\pk$ let $\T(a)$ denote the set of integral
weights which are images of $a$ under the action of $\T$,
  \be  \T(a):= \{ a+t \mid t\in\T \} \,. \ee

The key idea of the proof is to analyse the quantity
  \be  \Ss ab:= {\cal N}^{-1} \el^{-r}\! \sumt ca\, \sumt db \Sm{\elm c}d
  \labl s
for $a,b\in\pk$ ($\cal N$ is the normalization factor in the
Kac\hy Peterson formula \erf{kp}).
Because of the shift $t$ between $a$ and $c\in\T(a)$ (and between $b$ and $d$)
and because of the scaling by $\elm$, it is implicit in \erf s that we
consider the quantity $\Sm ab$ for weights which
are not necessarily integrable nor even integral; as already pointed out
in section 2, this is possible because
we are free to regard \erf{kp} as the {\it definition\/} of $\Sm ab$.

Using the simple fact that the finite Weyl group $W$ (in contrast to $\hat W$)
consists of linear maps and hence commutes with scalings,
it follows that for any pair $a,\;b$ of integral weights we can write
  \be  \begin{array}{l}  {\cal N}^{-1} \dstyle \sumt cb \Sm{\elm a}c
  \equiv \sumtT \sumww \signw\, \exp[-\pkg\, (\elm\,w(a),b+t) ]
  \\{}\\[-3.1 mm] \hsp{7.6} 
  = \dstyle\sumww \signw\, \exp[-\pkg\, \elm\,(w(a),b) ]
    \cdot \sumtt \exp[-\pkl\, (w(a),s) ]  \, \end{array}\labl{s1}
with
  \be  \TT := \kg^{-1}\,\T = \{ s\in\lv \mid 0\leq s_i<\el\ \,\forall \,
  i=1,2,...\,,r \} \,. \ee
Now for any fixed $w\in W$ we have
  \be  \begin{array}{l} \dstyle \sumtt \exp[-\pkl\, (w(a),s) ]
  \equiv  \sumtt \prod_{j=1}^r \exp[-\pkl\, (w(a))^j s_j ] \\{} \\[-3.1mm]
  \hsp{9.7}  
  = \dstyle\prod_{j=1}^r \llb \sum_{{\rm s}=0}^{\el-1}
   \exp[-2\pi\ii\, {\rm s}\, (w(a))^j/\el ] \lrb \\{} \\[-2.5mm]
  \hsp{9.7}  
  = \dprod_{j=1}^r \llb \el\,\deltal{(w(a))^j}0 \lrb
  \,,  \end{array}\labl2
where
  \be  \deltal pq:= \left\{ \begin{array}{ll} 1 & {\rm if}\ p=q\;\mod\el\,,
  \\[1mm] 0 & {\rm else}\,. \end{array}\right. \labl{19}
Next we use the elementary property of the Weyl group that Weyl
transformations map the weight {\it lattice\/} onto itself, i.e.\ that
for all $w\in W$ we have
  $    a^i\in\zett \mbox{ for all } i=1,2,...\,,r$ iff $
  (w(a))^i\in\zett \mbox{ for all } i=1,2,...\,,r$.
Then after defining
  \be  \deltalr ab:= \left\{ \begin{array}{ll} 1 & {\rm if}\ a^i=b^i\;\mod\el
  \ {\rm\ for\ all}\ i=1,2,...\,,r \,,
  \\[1mm] 0 & {\rm else}\,, \end{array}\right. \ee
analogously to \erf{19}, the formula \erf2 can be rewritten as
  \be  \sumts \exp[-\pkl\, (w(a),s) ] = \elr\cdot \prod_{j=1}^r \deltal{a^j}0
  = \elr\, \deltalr{a}0 = \elr\, \deltar{\elm a}0  \,. \labl3

Now from \erf3 it follows that the sum over $s\in\TT$
in \erf{s1} either vanishes or else just amounts to a factor of $\elr$;
hence when performing the corresponding sum in the quantity $\Ss ab$ we obtain
  \be  \Ss ab = {\cal N}^{-1} \!\!\sumt ca \deltar{\elm c}0 \, \Sm{\elm c}b
  = \sumww \signw \!\sumt ca \deltar{\elm c}0
  \exp[-\pkg\, \elm\,(w(c),b) ] \,. \labl{s2}

Next we show that any weight $\elm c$ which appears in the sum in \erf{s2} and
which
yields a non-zero contribution lies in fact on the Weyl orbit of a unique
element $d\in\Sig(a)$. To see this, we first notice that due to the projection
$\deltar{\elm c}0$ the relevant weights $\elm c$ are integral; moreover,
without loss of generality we can assume that they do
not lie on the boundary of any affine Weyl chamber, because otherwise
their contribution vanishes, $\Sm{\elm c}b=0$.
{}From the fact that $\hat W$ acts transitively and freely
on the interior of the chambers, it
then follows that there exists a unique $\wh\in\hat W$ such that the weight
$d:=\wh(\elm c)$ is integrable. Separating the translation part of $\wh$
from its finite Weyl group part, and inserting the explicit form
$c=a+\kg s$, $s\in\TT$, we arrive at the relation
  \be  d = w(\elm(a+\kg s))+\kg t=\elm\,\llb w(a) + \kg [w(s)+\el t] \lrb
  \labl d
for a unique $d\in\pk$ as well as for a unique $w\in W$ and a unique $t\in\lv$.
Now the weight $w(s)+\el t$ lies again in the coroot lattice \lv,
and hence by comparison with \erf{wla} we see that
\erf d indeed states that $d\in\Sig(a)$, which proves our claim.

Conversely, on the $W$-orbit of any $d\in\Sig(a)$ there is a
unique weight $\elm c$ with $c\in\T(a)$ which appears in \erf{s2} and
yields a non-zero contribution.
To check this, consider any fixed $d\in\Sig(a)$. Then
$a = \wh_d(\el d) = \el\,w_d(d)+\kg t_d$
for some $w_d\in W$ and some $t_d\in\lv$, or in other words
  $   \el d =\wh_d^{-1}(a) = w_d^{-1}(a) +\kg t'_d $
for some $t'_d\in\lv$. As $t'_d$ can be uniquely decomposed
as $t'_d=s'+\ell t'$ with $s'\in\TT$ and $t'\in\lv$, this means that to $d$
there
are associated an element $w_d$ of $W$ and weights $s'\in\TT,\ t'\in\lv$ such
that
  \be  \el d =w_d^{-1}(a) +\kg(s'+\el t')
  =w_d^{-1}(a +\kg w_d(s')) +\el\kg t' \,.  \labl{eld}
Now $w_d(s')$ lies in \lv\ so that it can be uniquely decomposed as
$w_d(s')=s+\el\,w_d(t'')$ with $s\in\TT$ and $t''\in\lv$; thus we can write
\erf{eld} as
  \be  \el d =w_d^{-1}(a +\kg s) +\el\kg(t'+t'') \,.  \labl{elD}
Since $t'+t''\in\lv$, we conclude that to any $d\in\Sig(a)$ there exists an
element $c\in\T(a)$, namely $c=a+\kg s$ with $s$ as constructed above,
such that $c$ lies on the $\hat W$-orbit of $\el d$.

Moreover, this element $c\in\T(a)$ is unique. Namely, according to
\erf{elD} $c$ not just lies on the $\hat W$-orbit of $\el d$,
but in fact already on the orbit of $\el d$ with respect to
the finite index subgroup $\wl\subset\hat W$. Therefore, assuming that both
$c=a+\kg s\in\T(a)$ and $c'=a+\kg s'\in\T(a)$ have this property, it follows
that there exists some $u\in\wl$ such that $c'=u(c)$. On the other hand, we
have of course $c'=w(c)$ with $w\in\hat W$ the translation
$w=(s'-s)\kg$. As $\hat W$ acts freely on the interior of the affine Weyl
chambers, it follows that $u=w$; since $u\in\wl$, we thus need in particular
$w\in\wl$ as well, which however implies that $w=\id$,\, i.e.\ $s=s'$.

Summarizing, we have proven that the weights $\elm c$ which
yield non-zero contributions to the sum in \erf{s2}
are in one-to-one correspondence with the elements of $\Sig(a)$,
with the explicit relationship given by \erf{elD}.
Consequently we can replace the summation over $\T(a)$ together with
the projection $\deltar {\elm c}0$ in \erf{s2} by a summation over $\Sig(a)$.
Furthermore the contribution from $\el\kg\lv$ to \erf{elD} can be
suppressed because upon insertion into \erf{s2} it just amounts to
trivial factors of unity. We then arrive at
  \be  \begin{array}{l}
  \Ss ab= \dstyle \sumww\signw \sum_{d\in\Sig(a)} \exp[-\pkg\,(w\circ
w_d(d),b)]
  \\ {} \\[-2 mm] \hsp{1.5} 
  = \dstyle \sum_{d\in\Sig(a)} \sign(w_d) \sumww\signw
  \exp[-\pkg\,(w(d),b)] = \sum_{d\in\Sig(a)} \epsl(d) \Sm db \,. \end{array}
\labl6
Thus we can conclude that the quantity $\Ss ab$ defined in \erf s coincides
with the expression on the left hand side of the sum rule \erf{1a}.

The remaining step in the proof of \erf{1a}, and hence of \erf1, is now
immediate. We just have to show that $\Ss ab$ coincides with the right hand
side of \erf{1a} as well. Because of the symmetry of $S$ this is equivalent
to showing that $\Ss\cdot\cdot$ is symmetric, too,
  \be  \Ss ba =\Ss ab \,. \labl y
Now the latter equality is an immediate consequence of \erf{slab} applied to
$\elm$:
  \be  \Ss ba= {\cal N}^{-1} \el^{-r} \sumt cb\, \sumt da \Sm{\elm c}d
  = {\cal N}^{-1} \el^{-r} \sumt da\, \sumt cb \Sm c{\elm d}  =\Ss ab \,. \ee
This concludes the proof of \erf1.

\sect{Quasi-Galois modular invariants}

To apply the result \erf1, consider for a given \qg\ scaling $\el$ the
matrix $\Pi$ with entries in $\{0,\pm1\}$ that describes the mapping induced by
the scaling on the primary fields, i.e.
  \be  \Pi_{a,b}\equiv \Pi_{a,b}^{(\el)} :
  = \epsl(a)\, \delta_{b,\sigmA a} \,. \labl{Pi}
As a consequence of \erf{01} one has
  \be  (\Pi S)_{a,b} = \epsl(a)\, \Sm{\sigmA a}b = \epsl(b)\,
  \Sm a{\sigmA b} = (S\Pi^t)_{a,b} \,,  \ee
while the sum rule \erf1 implies
  \be  (\Pi^t S)_{a,b} = \sump c \epsl(c)\, \delta_{a,\sigmA c} \Sm cb
  = \sump c \epsl(c)\, \delta_{b,\sigmA c} \Sm ac = (S\Pi)_{a,b} \,. \ee
Combining these results, it follows that the matrix
  \be  Z^{(\el)}:=\Pi+\Pi^t  \labl Z
commutes with the modular matrix $S$,
  \be  [Z^{(\el)},S] = 0 \,. \ee
Typically the \smat\ invariant $Z^{(\el)}$ obtained this way is not positive,
nor does it commute with
$T$. This pattern already arises for ordinary Galois scalings. However, just as
in the Galois case \cite{fgss,fusS}, it is still possible to construct physical
modular invariants, because one can get rid of the minus signs and achieve
$T$-invariance by suitably adding up various invariants of the type above and
possibly combining with other methods such as simple currents.
Note that in the invariant \erf Z typically some of the fields are projected
out, and hence when using \qg\ transformations it is in fact easier to obtain
$T$-invariance than in the Galois case.

To give an example for a matrix that commutes with the \smat\ and that is
obtained
by the above prescription, let us consider the scaling $\el=3$ for the
$A_1$ \wzwt\ at height $h=6$. In terms of non-shifted highest weights, this
scaling maps $\Lambda=0$ and $\Lambda=4$ with a positive sign $\epsl$
on $\Lambda =2$, the weight $\Lambda=2$ with a negative sign on itself,
and the weights $\Lambda =1,\, 3$ on the boundary $\BB$. Thus the matrix
$Z^{(3)}$ defined by \erf Z reads
  \small  \be
  \mbox{{\normalsize $Z^{(3)} =$}}
\left( \begin {array}{rrrrr} 0&0& 1&0&0\\\noalign{\smallskip}0&0&0&0&0
\\\noalign{\smallskip} 1&0& -2&0& 1
\\\noalign{\smallskip}0&0&0&0&0\\\noalign{\smallskip}0&0& 1&0&0
\end {array}\right) . \labl{qex}
\normalsize
While this matrix has negative entries and is hence unphysical, the combination
  \be  \hat Z = (Z^{(3)})^2 +2 Z^{(3)} \labl{qeX}
is a physical invariant, namely the $D$-type invariant of the height 6 $A_1$
theory. As the number of primary fields is rapidly increasing with the rank and
level, most applications of our prescription which lead to physical
invariants involve rather complex expressions; therefore we will not display
more complicated examples explicitly.

Actually the invariant \erf{qeX} can also be obtained from genuine Galois
transformations \cite{fusS}. An example for a
physical modular invariant which cannot be explained that way,
but which is obtainable as a linear combination of quasi-Galois invariants
is the exceptional $E_7$-type invariant of $A_1$ at level 16.
However, the concrete expression is rather lengthy so that
we refrain from presenting it here. As we shall see later, also for the
$E_7$-type invariant there exists a close relation to the matrix $Z^{(3)}$
displayed in \erf{qex} even though they are invariants at different heights.

\sect{\smat\ invariants: increasing and lowering the height}

In this section we consider the special case where the scaling factor
$\el\in\zetplus$ is a divisor of the height; to simplify notation, we will
make this explicit by denoting the height of
the theory to which the scaling is applied by $\lkg$. As we will see, in this
situation there exist intimate relations between the \wzwts\
at height $\lkg$ and at height $\kg$.
 \futnote{We are grateful to T.\ Gannon for remarks that triggered
the work presented in this section.}
 As we are now dealing with weights at two distinct heights, we find
it convenient to denote the elements of $\pk$ by lower case
and the elements of $\plh$ by upper case roman letters, respectively.
Similarly, we use the capital letter `\,$\SM$\,' for the \smat\ of
the height $\lkg$ theory and the symbol `\,$\sm$\,' for the \smat\
of the height $\kg$ theory.

Before describing the relationship between height $\kg$ and height $\lkg$
theories, let us first prove another new symmetry property of the \smat: if the
height is divisible by $\el$, then for any $B\in\plh$ the signed \smat\
elements
  \be  \epsl(C)\cdot\Smt{\el a}C  \labl{Slac}
are identical for all $C\in\Sig(B)$. To check this statement, take any fixed
$B\in\plh$ and any $C\in\Sig(B)$. Then considering
weights of the form $A=\el a$ with $a\in\pk$,
and using the fact that $\sigmA C=w_C(\el C)+\lkg\,t_C$ with $w_C\in W$ and
$t_C\in\lv$, as well as $\epsl(C)=\signwc$, we find
  \be  \begin{array}{ll}  \Smt{\el a}C= {\cal N} \dstyle \sumww\signw
  \exp[-\plkg\,(w(\el a),\elm w_C^{-1}(B)+\kg t_C')] \\ {}\\[-2 mm] \hsp{2.1}
  = {\cal N} \dstyle \sumww\signw \exp[-\pkg\,(w_Cw(a),\elm B)]
  \\ {}\\[-2 mm] \hsp{2.1}
  = \signwc\cdot {\cal N}\! \dstyle \sumww\signw \exp[-\pkg\,(w(a),\elm B)] \,.
  \end{array} \labl{slac}
The only dependence of the right hand side on the weight $C$ is thus via the
sign $\epsl(C)\equiv\signwc$, and hence we have established the symmetry
\erf{Slac}.
\smallskip

The primary WZW fields $\fI a$ and $\Fi A$ which are associated to the
weights in $\pk$ and in $\plh$, respectively,
can be viewed as the generators of the fusion rings \rh\ and \rlh\ of
the height $\kg$ and height $\lkg$ \wzwts, respectively.
Let us introduce the mappings
  \be  \begin{array}{ll}  \P: & \rlh\to\rh \\[2.2mm] & \  \Fi A \mapsto \P(\Fi
A)=
  \dstyle \sump b \P^{}_{A,b}\,\fI b \,, \qquad \P_{A,b}:= \epsl(A)\,
  \delta_{\sigmA A,\el b}  \end{array} \labl P
and
  \be  \begin{array}{ll}  \D: & \,\rh\to\rlh \\[2.2mm] & \ \fI a \mapsto \D(\fI
a)=
  \dstyle \sumP B \D^{}_{a,B}\,\Fi B \,, \qquad \D_{a,B}:=
  \delta_{\el a, B}  \, \end{array} \mbox{~~~~} \labl D
between these two fusion rings. Note that because of
  \be  \elm\sigmA A=\elm\,(w_A(\el A)+\lkg\,t_A)=w_A(A)+\kg\,t_A \labl{esa}
with $w_A\in W$ and $t_A\in\lv$ for any $A\in\plh$, the weight $\elm\sigmA A$
is
integral and either an element of $\pk$ or else on the boundary of an affine
Weyl
chamber at height $\kg$. Also, $\P_{b,b}=1$ (here the first label $b$ is
to be considered as an element of $\plh$) which shows that
the map $\P$ is always non-zero.

The relation \erf{esa} implies that there is a close connection, which
will prove to be useful later on, between
the conformal dimensions $\Delta$ mod \zett\ of all those fields which belong
to the same pre-image under the map $\sigmA$. Namely, from the definition
$\Delta_a=[(a,a)-(\rho,\rho)]/2\kg$ of the conformal dimensions at height
$\kg$ (and the
fact that any Weyl group element $w\in W$ is an isometry), it follows that
  \be  \begin{array}{l} \el\,(\Delta_b-\Delta_c) = (2\kg\el)^{-1}\,
  [(a+\kg t_b,a+\kg t_b)-(a+\kg t_c,a+\kg t_c)] \\[2.2 mm] \hsp{5.5}
  =\elm\,(a,t_b-t_c)+\half\kg\elm[(t_b,t_b)-\half(t_c,t_c)] \,
  \end{array} \labl{Ddl}
modulo \zett\
\,for $b,c\in\Sig(a)$. Since $t_b,\;t_c\in\lv$, we have $(a,t_b)\in\zett$,
$(t_b,t_b)\in2\zett$, and analogously for $t_c$, and hence the right hand sight
of \erf{Ddl} is an integral multiple of $\elm$. If in addition the
height is divisible by $\el$, then according to \erf{esa} this is also true
for the Dynkin components of any $a$ for which $\Sig(a)$ is non-empty, and
hence in this case the right hand side is in fact an integer, so that
  $    \Delta_b-\Delta_c \in \elm\zett$ for $h=\el h'$ and $b,c\in\Sig(a)$.
In the notation appropriate to the height $\lkg$ theory we thus
have, for all $A\in\plh$,
  \be  \Delta_B-\Delta_C \in \elm\,\zett \qquad {\rm for}\ \;
  B,C\in\Sig(A) \,.  \labl{ddl}

The relevance of the maps $\P$ and $\D$ that we introduced in \erf P and
\erf D comes from the fact that they provide direct relations between the
two modular matrices $\SM$ and $\sm$. Namely, we find
  \begin{eqnarray}  && \SM\,\D^t = \el^{-1/2}\,\P\,\sm \, \label i \\[2.2mm]
  && \P^t\,\SM = \elre\, \sm\,D \,.  \label{ii} \end{eqnarray}
Equivalently, by taking the transpose, we can write these identities as
  \begin{eqnarray}  && D\,\SM = \el^{-1/2}\,\sm\,\P^t  \, \label{iii} \\[2.2mm]
  && \SM\,\P = \elre\, D^t\,\sm \,.  \label{iv} \end{eqnarray}
The proof of of these relations is again rather technical; the reader who
wishes
to skip it may proceed directly to the paragraph after \erf{pts4}.
\medskip

To prove \erf i, we first separate the height-independent part of
the normalization factor $\cal N$ in the Kac\hy Peterson formula \erf{kp} from
the rest,
  \be  {\cal N} \equiv {\cal N}_{(\kg)} = \ii^{(d-r)/2}\,|L^{\rm w}/\lv|^{-1/2}
  \,\kg^{-1/2} =: \kg^{-1/2}\,\np \,.  \ee
Then we compute
  \be  \begin{array}{l}
  (\SM\,D^t)_{A,b}= \Smt A{\el b} = (\lkg)^{-1/2}\,\np \dstyle \sumww\signw
  \exp[ -\plkg\,(w(A),\el b)] \\ {} \\[-2mm] \hsp{3.8}
  = (\lkg)^{-1/2}\,\np \dstyle \sumww\signw \exp[ -\pkg\,(w(A),b)]  \end{array}
  \labl{sdt}
and, once again making use of $\sigmA A=w_A(\el A)+\lkg\,t_A$ with
$w_A\in W$ and $t_A\in\lv$, and of $\epsl(A)=\signwa$,
  \be  \begin{array}{l}
  (P\,\sm)_{A,b}= \epsl(A)\,\sM{\elm\sigmA A}b = \kg^{-1/2}\,\np \,\signwa
\dstyle
  \sumww\signw \exp[ -\pkg\,(w(w_A(A)+\kg t_A),b)] \\ {} \\[-2mm] \hsp{3.17}
  = \kg^{-1/2}\,\np \dstyle \sumww \signw \exp[ -\pkg\,(w(A),b)] \,.
\end{array}
  \labl{ps}
Comparing \erf{sdt} and \erf{ps}, we obtain \erf i.

The proof of \erf{ii} requires a bit more work. We first observe that
  \be  (\sm\,D)_{a,B}= \sumpc \sM ac\,\delta_{\el c,B}= \left\{
\begin{array}{ll}
  \sM a{\elm B} & {\rm if}\ B=\el b\ \;{\rm for\ some}\ b\in\pk \,,
  \\[1.2mm] 0 & {\rm else}\,. \end{array} \right.  \labl{sd}
Second, with the help of the sum rule \erf1, we obtain
  \be  \begin{array}{l}  (P^t\,\SM)_{a,B}= \dstyle \sumP C \epsl(C)\,
  \delta_{\el a,\sigmA C}\,\Smt CB \equiv
  \sum_{C\in\Sig(\el a)} \!\!\epsl(C)\, \Smt CB \\ {}\\[-2mm] \hsp{3.77}
  = \dstyle \sum_{C\in\Sig(B)} \epsl(C)\, \Smt{\el a}C  \,.  \end{array}
\labl{pts1}
Next we notice that according to the property \erf{Slac} of the \smat\ the
terms
in the sum over $C$ on the right hand side of \erf{pts1} are actually
independent
of $C$ (and are of the specific form obtained in \erf{slac}), so that
  \be  (P^t\,\SM)_{a,B} = |\Sig(B)| \cdot (\lkg)^{-1/2}\,\np
  \sumww\signw \exp[-\pkg\,(w(a),\elm B)] \,. \labl{pts2}
Now if $B\ne\el b$ for all $b\in\pk$, then according to \erf{esa} the set
$\Sig(B)$ is empty. On the other hand,
for $B=\el b$ with $b\in\pk$, \erf{pts2} can be rewritten as
  \be  (P^t\,\SM)_{a,B} = |\Sig(B)| \cdot \el^{-1/2}\,\sM ab \,. \labl{pts3}
Moreover, in this case we have $|\Sig(B)|=\elr$ because for $B=\el b$ the
elements $C\in\Sig(B)$ are of the form
$C=w(b)+\kg t$ with $w\in W$ and $t\in\lv$, and furthermore the fundamental
affine Weyl chamber at height $\lkg$ consists of $\elr$ of the affine Weyl
chambers at height $\kg$, so that the orbit of $b\in\pk$ with respect to the
height $\kg$ affine Weyl group contains $\elr$ weights $b'$ which belong to
$\plh$,
and these weights $b'$ are precisely those which are of the form required for
$C\in\Sig(B)$.
Thus we can conclude that $(P^t\,\SM)_{a,B}$ vanishes unless
$B=\el b$ for some $b\in\pk$, in which case we have
  \be  (P^t\,\SM)_{a,B} = \elre\,\sM a{\elm B} \,. \labl{pts4}
Comparison of this result with \erf{sd} then completes the proof of \erf{ii}.
\medskip

We can now apply the results just proven to the construction of \smat\
invariants, both at height $\kg$ and at height $\lkg$.
Namely, assume first that the matrix $Z$ belongs to the commutant of the \smat\
of the height $\lkg$ theory, i.e. that
  \be  [Z,\SM] = 0  \,. \labl{Zs}
Further, define
  \be  \zt:= \P^t\,Z\,\D^t+\D\,Z\,\P \,.  \labl{zt}
Explicitly, we have
  \be  \zt_{a,b}= \sum_{A\in\Sig(\el a)} \epsl(A)\, Z_{A,\el b} +
  \sum_{B\in\Sig(\el b)} \epsl(B)\, Z_{\el a,B} \,.  \ee
Using \erf{Zs} as well as the relations \erf i -- \erf{iv} proven above, we
can then derive that
  \be  \begin{array}{l}  \zt\,\sm = \P^t\,Z\,\D^t\,\sm+\D\,Z\,\P\,\sm
  = \el^{-r+1/2}\,\P^t\,Z\,\SM\,\P+\el^{1/2}\D\,Z\,\SM\,D^t  \\[1.9mm]
  \hsp{1.02} 
  = \el^{-r+1/2}\,\P^t\,\SM\,Z\,\P+\el^{1/2}\D\,\SM\,Z\,D^t
  = \sm\,\D\,Z\,\P + \sm\,\P^t\,Z\,\D^t = \sm\,\zt \,. \end{array}\ee

Similarly, let $z$ be an \smat\ invariant of the height $\kg$ theory,
  \be  [z,\sm] = 0  \,, \labl{zs}
and define
  \be  \Zt:= \D^t\,z\,\P^t+\P\,z\,\D \,. \mbox{~~} \labl{Zt}
Using the convention that $z_{a,b}=0$ whenever $a$ or $b$ is not in $\pk$,
the matrix elements of $\Zt$ read
  \be  \Zt_{A,B}= \epsl(A)\, z_{\elm\sigmA A,\elm B} +
  \epsl(B)\, z_{\elm A, \elm\sigmA B} \,. \labl{ZT}
By employing \erf{zs} and again \erf i -- \erf{iv}, we obtain
  \be  \begin{array}{l}  \Zt\,\SM = \D^t\,z\,\P^t\,\SM+\P\,z\,\D\,\SM
  = \elre\,\D^t\,z\,\sm\,\D+\el^{-1/2}\,\P\,z\,\sm\,P^t \\[2.2mm]
  \hsp{1.45} 
  = \elre\,\D^t\,\sm\,z\,\D+\el^{-1/2}\,\P\,\sm\,z\,P^t
  = \SM\,\P\,z\,\D +\SM\,\D^t\,z\,\P^t = \SM\,\Zt \,. \end{array}  \mbox{~~~}
\ee

We have thus proven the following remarkable facts: Given an \smat\ invariant
$Z$
at height $\lkg$, the formula \erf{zt} provides us with an \smat\ invariant
$\zt$
at height $\kg$,
  \be  [\zt,\sm]=0 \,; \ee
and conversely, given an \smat\ invariant $z$ at height $\kg$, the formula
\erf{Zt} defines an \smat\ invariant $\Zt$ at height $\lkg$,
  \be  [\Zt,\SM]=0 \,. \ee
Not surprisingly, the prescriptions \erf{zt} and \erf{Zt} do not respect
positivity, i.e.\ even if $Z$ (respectively $z$) is a positive invariant, this
needs not hold for $\zt$ ($\Zt$).

As an example, let us take for $Z$ the exceptional invariants of $A_1$
which occur all at heights a multiple of 6, namely for $h= 12, 18, 30$, and
obtain from them by \erf{zt} invariants of $A_1$ at height 6.
For $h=12$ and $h=30$ the prescription \erf{zt} yields the zero matrix.
More interesting is the $E_7$-type invariant at $h=18$; in this case
$\zt$ is precisely the quasi-Galois invariant \erf{qex} obtained in the
previous section.
\smallskip

Note that the maps \erf P and \erf D are related to the map $\Pi$
introduced in \erf{Pi} by $\Pi=\P\D$:
  \be  \Pi_{A,B}=\epsl(A)\,\delta_{B,\sigmA A}
  \equiv \sump c \epsl(A) \, \delta_{\el c,\sigmA A} \delta_{B,\el c}
  = \sump c \P_{A,c}D_{c,B} \,. \labl{ppd}
The prescription \erf{Zt} actually provides a generalization of
the \qg\ \smat\ invariant \erf Z. Namely,
according to \erf{ppd}, when considering the diagonal invariant $z=\one$,
\erf{Zt} yields
  \be  \Zt = \P\,\D + \D^t\,\P^t = \Pi + \Pi^t  \,,  \ee
i.e.\ reproduces the invariant \erf Z.
A still more special case is obtained by performing the scaling by
the factor $\el$ at height $\el\gv$. Then the smaller level is in fact zero,
so that there is a single primary field with shifted weight $a=\rho$, and hence
a
single nontrivial invariant $z_{a,b} = \delta_{a,\rho} \delta_{b,\rho}$.
In this situation, \erf{ZT} reads
  \be  \Zt_{A,B} =
  \delta_{A,\el\rho} \sum_{C\in\Siglr} \epsl(C)\,\delta_{B,C} +
  \delta_{B,\el\rho} \sum_{C\in\Siglr} \epsl(C)\,\delta_{A,C} \,. \labl x
In applications (see in particular section \ref{sce} below) it is often
not the matrix \erf x that is directly relevant, but rather the combination
  \be  \zh:= \Zt^2 - 2\epsl(\el\rho)\,\Zt  \labl{zh}
(compare the similar formula \erf{qeX}). The entries of \erf{zh} read
  \be  \zh_{A,B} = |\Siglrm|\, \delta_{A,\el\rho} \delta_{B,\el\rho}
  + \sum_{C,D\in\Siglrm} \!\! \epsl(C)\epsl(D)\,\delta_{A,C}\,\delta_{B,D}
  \,, \labl o
where
  \be  \Siglrm:=\Siglr\,\setminus \{\el\rho\}  \,.  \labl'
Note that in the invariant $\zh$ only fields belonging to $\Sig(\el\rho)$
get mixed; by \erf{ddl} this implies that $\zh$ is not only $S$-invariant,
but also invariant under $T^{\el}$.
It is also easily checked that $\zh^2=|\Siglrm|\,\zh$, so that by taking
powers of $\zh$ we cannot produce any new invariants.

We can also apply the constructions \erf{Zt} and \erf{zt} consecutively to a
height $\kg$ \smat\ invariant, or in the opposite order to a
height $\lkg$ invariant. The computation then involves the identities
$\P\D=\Pi$, $\D\D^t=\one$, $\P^t\P=\elr\one$, as well as
$\D\P=\pi$ and $\D^t\D=\Pr$ with
  \be  \pi_{a,b}:= \epsl(\el a)\,\delta_{\el b,\sigmA(\el a)}  \labl{pi}
and
  \be  \Pr_{A,B}:= \delta_{A,B}\cdot \sump b \delta_{A,\el b} \,. \labl{Pr}
We find
  \be  \ztt= 2\elr\,z + \pi z\pi + \pi^t z\pi^t  \labl{ztt}
and a similar formula for $\Ztt$.
The result \erf{ztt} means that whenever $z$ commutes with $\sm$, then so does
the matrix $ \pi z\pi + \pi^t z\pi^t$. Also note that in \erf{pi} the map
$\sigmA$
is the \qg\ transformation with scale factor $\el$ at height $\lkg$.
This implies that $\sigmA(\el a)=\el\,(w_{\el a}(\el a)+\kg t_{\el a})$, and
hence
the $\delta$-symbol in \erf{pi} imposes the constraint that the weight $b$ is
related to $a$ by a \qg\ transformation with the same scale factor $\el$, but
now
at height $\kg$. In other words, as already anticipated in the notation,
the map $\pi=DP$ implements the same \qg\ scaling for the height $\kg$ theory
as the map $\Pi=PD$ \erf{ppd} implements for the height $\lkg$ theory.

\sect{Conformal embeddings} \label{sce}

Conformal embeddings
are embeddings $\g\emb\h$ of untwisted affine Lie algebras for
which the irreducible highest weight modules possess finite branching rules.
The explicit form of these branching rules has been determined for various
cases (see e.g.\ \cite{kasa,kawa,walt2,chra2,hase,albi,vers2}), but
a general formula is not known, and there are still many
conformal embeddings for which all known methods are inapplicable.

The list of conformal embeddings \cite{scwa,babo} contains several infinite
series.
Here we are interested in a particular infinite series,
namely the embedding $\g_{\gv}\emb\SOD_1$, i.e.\ of \g\ at level $\gv$
(with \g\ an arbitrary
untwisted affine \lie) into \SOD, with $d\equiv{\rm dim}\,\gb$, at level one.
In terms of the horizontal algebras, the embedding is the one for which the
vector representation of \sod\ branches to the adjoint representation of the
smaller algebra \gb. Such embeddings are of particular interest because they
are connected with the `fermionization' of WZW models with level $\gv$,
which is due to the fact that \SOD\ can be written in terms of free fermions.
This
will play a r\^ole in the following.

The diagonal level one \SOD\ partition function is
  \be   {\calP}(\tau, \bar\tau) =|\chio|^2 + |\chiv|^2 +
  |\chis|^2 + |\chic|^2 \qquad \hbox{for $d$ even}   \labl{Diag}
and
  \be   {\calP}(\tau, \bar\tau) =|\chio|^2 + |\chiv|^2 +
  |\chis|^2  \qquad\qquad \hbox{for $d$ odd,}   \labl{DiaG}
where o,\,v,\,s and c refer to the singlet, vector, spinor, and conjugate
spinor
\rep\ of \sod, respectively.
Our objective is to write each of these characters in terms of
characters of \g\ at level $\gv$.

The branching rule for the \SOD\ spinor(s) is already known ($\!\!$\cite{kape},
see also \cite{gool,jf11}). Up to a multiplicity, they branch to a single
irreducible representation, namely the one whose (unshifted)
highest weight is the Weyl vector $\rho$. We will denote this irreducible
\rep\ by \rhO. The dimension of the analogous
irreducible representation of the horizontal algebra \gb\ is $2^{N_+ }$, where
$N_+=(d-r)/2$ is the number of positive roots (and $r$ is the rank of \gb);
hence
the multiplicity with which $\rhO$ is contained in the
\SOD\ spinors is $2^{{r/2}-1}$ if $d$ is even, and
$2^{{(r-1)/2}}$ if $d$ is odd. Then we can make the following ansatz for the
relation between level one \SOD\ and $\Kacg_{\gv}$ characters:
  \be  \chio = \sum_i \moi\, \calYi \,, \qquad
  \chiv = \sum_i \mvi\, \calYi \,, \qquad
  \chis = \chic = 2^{{r/2}-1} {\calY}_{\rho} \ee
for $d$ even, and
  \be  \chio = \sum_i \moi\, \calYi \,, \qquad
  \chiv = \sum_i \mvi\, \calYi \,, \qquad
  \chis = 2^{{(r-1)/2}} {\calY}_{\rho}  \mbox{~~~~~}\ee
for $d$ odd.
Here the sums are over all integrable $\Kacg_{\gv}$ representations,
i.e.\ over $P_{2\gv}(\g)$, and we labeled them
by their unshifted highest weights; $\mo$ and $\mv$ are non-negative
integral vectors in the space of all characters. The equality of the
decomposition of the two \SOD\ spinor characters for even $d$ implies that
these
representations will appear as a fixed point of order 2 in the $\Kacg_{\gv}$
modular invariant. Hence the invariant will have the form
  \be   \zcec= | \sum_i \moi\, \calYi|^2 + | \sum_i \mvi\, \calYi |^2 +
  2\cdot | 2^{{r/2}-1} {\calY}_{\rho}|^2 \labl{65}
for $d$ even, and
  \be   \zcec= | \sum_i \moi\, \calYi|^2 + | \sum_i \mvi\, \calYi |^2 +
  |  2^{{(r-1)/2}}{\calY}_{\rho}|^2  \mbox{~~} \labl{66}
for $d$ odd.

The identity and vector characters of \SOD\ branch to distinct $\Kacg_{\gv}$
characters, since the difference of conformal dimensions of identity and
vector is non-integral. As a consequence, the vectors $\mo$ and $\mv$ are
orthogonal. We will focus first on the cases where
also the spinor(s) have different conformal weights modulo integers than
identity and vector, which holds if $d \not = 0\; \mod 8$. Then by the
same argument the spinor(s) branch to different $\Kacg_{\gv}$ characters than
identity and vector characters, and hence we have $\mor=\mvr=0$.

This situation is covered by the following simple theorem.
Consider any $S$-invariant (such as \erf{65}, \erf{66})
that is a sum of squares, i.e.\ of the form
  \be  {\cal M}= \sum_{\Ell} N_{\Ell}\, | \sum_i \mpi\, \calX_{i} |^2
  \ . \ee
This can be written as $\sum_{i,j}{\calX}_i^{} M_{ij}^{} {\calX}^*_j$, where
$M$ is the matrix with entries
  \be  M_{ij}=\sum_{\Ell} N_{\Ell} \mpi\mpj  \;. \ee
Further, suppose that the vectors $\mpe$ are orthogonal,
  \be  \sum_i \mpi\,\mpI =R_{\Ell}  \delta_{\Ell\Ell'} \,. \ee
Let us also impose the physical requirement that there is a unique vacuum,
i.e.\ that $M$ satisfies $M_{00}=1$; then
among the vectors $\mpe$ there must be precisely one, conventionally
labeled by $\Ell=0$, which contains the identity character, i.e.\ we must
have $N_\OO=1$ and $\mOO=1$. Next consider the matrix $M^2$; it has entries
$(M^2)^{}_{ij}=\sum_{\Ell} N_{\Ell}^2 R_{\Ell} \mpi\,\mpj$; in
particular, $(M^2)_{\OO\OO}=R_\OO$. Thus the matrix $M^2 - R_\OO M$ has entries
$(M^2- R_\OO M)_{ij} = \sum_{\Ell} (N_{\Ell}^2 R_{\Ell}-N_{\Ell} R_\OO)\,
\mpi\,\mpj\,.$ Finally, the square $Z$ of the latter matrix has entries
  \be  Z_{ij} \equiv ( [M^2 - R_\OO M]^2)_{ij} =\sum_{\Ell} (N_{\Ell} R_{\Ell}
  - R_\OO)^2 N_{\Ell} R_{\Ell} \,\mpi\,\mpj \;. \labl{60}
This is a manifestly non-negative matrix, it obeys $Z_{\OO\OO}=0$, and
because it is a polynomial in $M$ it commutes with $S$. Thus $0=Z^{}_{\OO\OO}
= \sum_{i,j} S_{\OO i} Z_{ij} S_{\OO j} \geq 0$, with equality only
if $Z_{ij}=0$ for all $i$ and $j$; i.e.\ any such matrix must vanish.
By \erf{60}, the vanishing of $Z$ implies that for any $\Ell$ the sum rule
  \be  N_{\Ell}\sum_i (\mpi)^2 \equiv N_{\Ell} R_{\Ell}=R_{\OO} \labl{sr}
holds. This is equivalent to the property $M^2=R_\OO M$, so that $M$
is idempotent up to a normalization.

In the situation of our interest, these sum rules
gives useful information because we know $N_{\Ell}$ and $\mpe$
for the spinor characters. For even $d$, the spinors have $N=2$, and
hence \erf{sr} tells us that
  \be  R_\oo = N_\vv R_\vv = 2\cdot (2^{{r/2}-1})^2 = 2^{r-1} \,,\ \ \
  \labl{SumRuleA}
and for $d$ odd we get
  \be  R_\oo = N_\vv R_\vv = (2^{{(r-1)/2}})^2 = 2^{r-1} \,.\ \ \ \
  \labl{SumRuleB}
Since for $d\neq8\; \mod16$ the vector representation of level one \SOD\ has
different conformal dimension modulo integers than the other \rep s, we
have $N_\vv=1$. In all examples we know of the matrix $M$ has all entries
except the
spinor entries equal to $0$ or $1$, and in that case the sum rule \erf{sr}
tells us that the identity and the vector of \SOD\ each branch to $2^{r-1}$
different irreducible representations of the conformal subalgebra \g.

This is what one can say about these invariants by rather general arguments.
We will now discuss how one can conjecture the form
of these invariants (i.e.\ the form of the vectors $\mo$ and $\mv$)
by employing a quasi-Galois scaling by a factor 2.
Thus consider \g\ at height $h=2\gv$, and the \qg\ scaling $\el=2$.
Applying the prescription \erf{Zt}, we obtain the special case $\el=2$
of the \smat\ invariant \erf o. Using unshifted weights (in particular
$\Lambda=\rho$ in place of $a=2\rho$), \erf o reads
  \be  \zh_{\Lambda,\Lambda'} = |\SiglRm|\, \delta_{\Lambda,\rho}
  \delta_{\Lambda',\rho} + \sum_{\mu,\mu'\in\SiglRm} \!\! \epsz(\mu)\epsz(\mu')
  \,\delta_{\Lambda,\mu}\,\delta_{\Lambda',\mu'}  \,. \labl{zhl}
As it turns out, the sign $\epsz$ is not constant on $\Sigzr$, so that
(unlike in the, otherwise similar, situation of \erf{qex})
the invariant $\zh$ \erf{zhl} is not positive.
By the remark after \erf' it follows,
however, that it does commute with $T^2$.

Further, for all simple \gb\ except $\gb=A_r$ with $r$ even, we observe the
following. A certain number $K$ of representations with integer conformal
weight is mapped via the \qg\ transformation
to $\rhO$ with a positive sign; an equal number of representations
with half-integer conformal weight flows to $\rhO$ with a negative sign;
all other representations as well as $\rhO$ itself flow to the boundary.
(This has been checked explicitly for rank less than 9; the
continuation to higher rank is only a conjecture.) For $A_r$ with $r$ even,
there are two differences with respect to the foregoing.
First of all the numbers $K$ and $K'$ of fields with integral and half-integral
conformal weight, respectively, that flow
to $\rhO$ are different, and secondly $\rhO$ does not flow to the boundary, but
to itself. In this case $d=r(r+2)$, which is a multiple of $8$, implying that
the
\SOD\ spinor has integral or half-integral conformal weight. The sign
associated with the flow of $\rhO$ to itself is plus or minus for these two
cases respectively.

In matrix notation, we thus have $\Zt=\Pi+\Pi^t$, with
  \be \Pi = \left (\begin{array}{cccc} 0 & 0 & \vecv & 0 \Cr
  0 & 0 & -\vecv & 0 \Cr 0 & 0 & \epsz(\rho) & 0 \Cr 0 & 0 & 0 & 0 \end{array}
  \right )\ , \labl{PI}
for the matrix \erf{Zt} that underlies \erf{zh}, and hence
  \be  \zh= \left ( \begin{array}{cccc}
  \EE& -\EE & 0 & 0 \Cr -\EE & \EE & 0 & 0 \Cr
  0 & 0 & K+K' & 0 \Cr 0 & 0 & 0 & 0 \end{array} \right ) \;. \labl{FQG}
Here the third column/row corresponds to $\rhO$, the first one to
all $K$ fields with integral conformal weight which flow to $\rhO$ under
the \qg\ transformation, the second to the
$K'$ fields with half-integral weight flowing to $\rhO$, and the fourth
to all remaining fields. The symbol $\vecv$ stands for
a $K$, respectively $K'$, component vector with all entries equal to 1,
and $\EE\equiv\vecv\otimes\vecv^{\,t}$ denotes the matrix
of appropriate size (i.e., $K\times K$, $K\times K'$, $K'\times K$, and
$K'\times K'$, respectively) each of whose entries is equal to 1;
the 0's indicate matrices of zeroes of the proper size,
Thus in particular for all cases except $A_r$ with even rank, \erf{FQG}
can also be written as
  \be  \zh= \left ( \begin{array}{cccc}
  \EE& -\EE & 0 & 0 \Cr -\EE & \EE & 0 & 0 \Cr
  0 & 0 & 2K & 0 \Cr 0 & 0 & 0 & 0 \end{array} \right )  \labl{FromQG}
with all matrices $\EE$ of size $K\times K$.
Also recall that if $\rhO$ flows to the boundary, then $\epsz(\rho)=0$
so that the entry $\Pi_{\rho,\rho}$ of the matrix \erf{PI} vanishes.
Further, if $d$ is a multiple of 8, then not only the matrix \erf{FQG},
but also
  \be  \zh':=\zh+\epsz(\rho)\,\Zt= \left ( \begin{array}{cccc}
  \EE& -\EE & \epsz(\rho)\,\vecv & 0  \Cr -\EE & \EE & -\epsz(\rho)\,\vecv
  & 0 \Cr \epsz(\rho)\,\vecv ^{\,t}& -\epsz(\rho)\,\vecv^{\,t}
  & K\!+\!K'\!+\!2\epsz^2(\rho) & 0
  \Cr 0 & 0 & 0 & 0 \end{array} \right )  \labl{FQG'}
commutes with both $S$ and $T^2$.

These results can be related to the expected conformal embedding in the
following way. Consider first the case of
even $d$. The diagonal \SOD\ invariant can be written in terms of Jacobi
theta functions and the Dedekind eta function, using
  \be  \begin{array}{ll}
  \chio = \half \eta^{-\nn} (\theta_3^\nn + \theta_4^\nn ) \,, &
  \chiv = \half \eta^{-\nn} (\theta_3^\nn - \theta_4^\nn  ) \\[2.7 mm]
  \chis = \half \eta^{-\nn} (\theta_2^\nn + \ii^\nn\theta_1^\nn) \,,\mbox{~~} &
  \chic = \half \eta^{-\nn} (\theta_2^\nn - \ii^\nn \theta_1^\nn) \
,\end{array}
  \labl{69}
where the arguments $\tau$ and z are suppressed.
We are only considering Virasoro specialized characters here, i.e.\
these functions are in fact $\theta_i({\rm z}=0,\tau)$. Since $\theta_1({\rm z}
=0,\tau)=0$, in this setting the partition function \erf{Diag} reads
  \be  {\calP} = \half \, |\eta|^{-\nN}
  \left[\, |\theta_3|^{\nN} + |\theta_4|^{\nN} + |\theta_2|^{\nN} \,\right]
  \,. \ee
This is modular invariant because $S$ interchanges $\theta_4$ and $\theta_2$,
while $T$ interchanges $\theta_4$ and $\theta_2$, and all overall factors
cancel.

This diagonal partition function is however not the one we obtain from
quasi-Galois transformations. Using the modular transformation properties
of the $\theta$-functions one can write down another partition function
that is only invariant under $S$ and $T^2$:
  \be  {\caLP} =  |\eta|^{-\nn} \left[\, |\theta_4|^{\nN} + |\theta_2|^{\nN}
  \,\right] \,. \mbox{~~} \labl{Pprime}
We can re-express this in terms of the \SOD\ characters \erf{69} to obtain
  \be  {\caLP} = | \chio  - \chiv|^2 + | \chis + \chic |^2 \;. \labl{NonDiag}
(The normalization of \erf{Pprime} was chosen to make the square of the
identity character appear exactly once.)
Both the diagonal modular invariant \erf{Diag} and the partition function
\erf{NonDiag} contain more information than one strictly
gets from specialized characters; one may check explicitly that both
are formally  $S$-invariant if the spinor characters are distributed
symmetrically, as indicated.

If we write the matrix $M$ corresponding to
\erf{NonDiag} in terms of \g-representations we get
  \be  \left (\begin{array}{cccc}
  \EE_{\oo\oo} & -\EE_{\oo\vv} & 0 & 0 \Cr -\EE_{\vv\oo} & \EE_{\vv\vv}  & 0
  & 0 \Cr 0 & 0 & 2^r & 0 \Cr 0 & 0 & 0 & 0 \end{array} \right ) \ ,
\labl{SubMat}
where $(\EE_{\Ell\Ell'})_{ij}=\mpi\,\mpJ$.
This can be identified with \erf{FromQG} provided that
  \be  \EE=\EE_{\oo\oo}=\EE_{\oo\vv}=\EE_{\vv\oo}=\EE_{\vv\vv} \ , \ee
or in other words, that $\mov=\mvv = \vecv$. Although we cannot
prove that this identification is correct, we have a direct
consistency check. Namely, we find that $K=2^{r-1}$, and hence that both $\mo$
and
$\mv$ have $2^{r-1}$ components, each equal to 1. Hence they do satisfy the
sum rule \erf{SumRuleA}, so this rather nontrivial requirement for the matrix
  \be   \zce:=\left (\begin{array}{cccc}
  \EE & 0 & 0 & 0 \Cr 0 & \EE & 0 & 0 \Cr
  0 & 0 & 2^{r-1} & 0 \Cr 0 & 0 & 0 & 0 \end{array} \right ) \  \labl{ConfEmb}
to commute with $S$ is fulfilled. The matrix \erf{ConfEmb}
is the conjectured modular invariant
corresponding to the conformal embedding. Unfortunately the
quasi-Galois symmetries allow us only to conclude that \erf{FromQG}
commutes with $S$ and $T^2$, but the step from \erf{FromQG} to
\erf{ConfEmb} does not follow from any symmetry we know.

If $d$ is a multiple of 8, then the above argument has to be slightly extended.
Since in this case both \erf{FQG} and \erf{FQG'} are $S$-$T^2$-invariants,
we have in addition to \erf{ConfEmb} another matrix $\zCe$,
and hence any physical linear combination
  \be  Z(u,v):= u\,\zce + v\,\zCe  \,,  \labl{uv}
as candidates for the conformal embedding invariant.
Explicitly, the matrix $\zCe$ reads
  \be   \zCe:=\left (\begin{array}{cccc}
  \EE & 0 & \vecv & 0 \Cr 0 & \EE & 0 & 0 \Cr
  \vecv^{\,t} & 0 & 2^{r-1}\!+\epsz^2(\rho) & 0 \Cr 0 & 0 & 0 & 0 \end{array}
  \right ) \  \labl{z0}
for $d=0\;\mod\,16$ and
  \be   \zCe:=\left (\begin{array}{cccc}
  \EE & 0 & 0 & 0 \Cr 0 & \EE & \vecv & 0 \Cr
  0 & \vecv^{\,t} & 2^{r-1}\!+\epsz^2(\rho) & 0 \Cr 0 & 0 & 0 & 0 \end{array}
  \right ) \  \labl{z8}
for $d=8\;\mod\,16$, respectively. Fortunately, it is easy to eliminate
all but one of the candidates, namely by imposing the `\qdim' sum rule
  \be   \half = {(S^{}_{{\rm so}(d)})}^{}_{\OO,\OO} = \sum _i
  {(S^{}_{\g})}^{}_{\OO,i} \labl{kawa}
(here the summation is over all fields that are combined with the
identity field). Inserting the ansatz \erf{uv}, we find that
for the case of $A_r$ with even $r$, this yields the unique
solution $u=0,\,v=1$, so that \erf{z0}, respectively
\erf{z8}, is the correct candidate (and we also have $\epsz^2(\rho)=1$).
In contrast, for all other cases where $d$ is a multiple of 8 (such as
\gb=$E_8$), the unique solution is given by $u=1,\,v=0$, i.e.\
only \erf{ConfEmb} survives the constraint \erf{kawa}. Thus
in all cases except $A_r$ with $r$ even the
situation is the same as in the general case where $d$ is not divisible by 8.

For odd $d$ the use of theta functions is somewhat awkward, but
it suffices to observe that the matrix
  \be  M=\left( \begin{array}{rrr}
  1 & -1 & 0 \\[.5mm] -1 & 1 & 0 \\[.5mm] 0 & 0 & 2 \end{array}\right) \labl;
commutes with the \smat
  \be   S^{}_{{\rm so}(d)} = \half \left( \begin{array}{ccc}
  1 & 1 & \sqrt2 \\[1.1mm] 1 & 1 & -\sqrt2 \\[1.1mm]
  \sqrt2 & -\sqrt2 & 0 \end{array}\right) \ee
Written in terms of \g-characters, \erf; becomes
identical to \erf{SubMat}, and the rest of the argument is the same.

\medskip
In the notation of \erf{zhl}, the conjectured conformal embedding
invariant \erf{ConfEmb} reads
  \be  \zcE_{\Lambda,\Lambda'} = 2^{r-1}\,\delta_{\Lambda,\rho}
  \delta_{\Lambda',\rho}
  + \sum_\SiglrMP \delta_{\Lambda,\mu} \delta_{\Lambda',\mu'}
  + \sum_\SiglrMM \delta_{\Lambda,\mu} \delta_{\Lambda',\mu'}  \,, \labl{zce}
while \erf{z0} and \erf{z8} with $\epsz(\rho)=\pm1$ can be summarized as
  \be  \zCE_{\Lambda,\Lambda'} = (2^{r-1}+1)\,\delta_{\Lambda,\rho}
  \delta_{\Lambda',\rho}
  + \sum_\SiglrMP \delta_{\Lambda,\mu} \delta_{\Lambda',\mu'}
  + \sum_\SiglrMM \delta_{\Lambda,\mu} \delta_{\Lambda',\mu'}  \,. \labl{zce'}
Accordingly, the conjectured branching rules read
  \be  \chio=\sum_\SIglrmp \chi_\mu^{} \,, \qquad\quad
  \chiv=\sum_\SIglrmm \chi_\mu^{} \,. \labl{BR}
Note that in the summations the weight $\mu=\rho$ does not contribute
except for $A_r$ with even $r$, in which case it contributes to
$\chio$ (if $d\equiv r(r+2)=0\;\mod\,16$) and to $\chiv$
(if $d=8\;\mod\,16$), respectively.
\medskip

In addition to the consistency check already mentioned, our conjecture also
passes several other non-trivial tests:
First, the matrix \erf{zce} commutes with $T$.
Second, by inspection one verifies that the correct number ${\rm dim}(\sod)-
{\rm dim}(\gb)=d(d-3)/2$ of spin one currents are combined with the
identity field. Third, again by inspection one checks that the `\qdim' sum
rule \erf{kawa} is satisfied also for $d$ not a multiple of 8, where the
sum rule was not used in our argument.
And finally, in the few cases where the branching had already been
determined before, such as for $\gb=G_2$ \cite{chra2}, we reproduce the
known result.
To us these observations make it almost inevitable that the branching rules
for the embedding $\g_{\gv}\emb\SOD_1$ are indeed given by \erf{BR}.

\medskip
Let us also present some examples for the conjectured invariants. The
most interesting cases are those with exceptional \gb. We will display
the result for the algebras $\gb=F_4$ and $\gb=E_6$ (in the $E_7$ and
$E_8$ cases the invariants require too much space, therefore they will
be presented elsewhere \cite{sche7}). The primary fields are again labeled
by their unshifted highest weights. We find
\small
\def\Cr{\\[1.2mm]} \def\CR{\\[2.6mm]}
  \be  \begin{array}{ll} \mbox{{\normalsize $\zcec(F_{4,9}) =$ }}
 &        |\, (0,0,0,0) + (0,0,1,6) + (0,0,2,1) + (0,1,0,0)       \Cr
 &    \quad + (0,1,1,2) + (0,3,0,0) + (1,0,0,5) + (1,1,0,4) \,|^2 \CR
 &\!\!+\, |\, (0,0,0,7) + (0,0,2,0) + (0,0,3,0) + (0,1,0,3)       \Cr
 &    \quad + (0,1,0,6) + (0,2,0,2) + (1,0,0,0) + (1,0,1,4) \,|^2 \CR
 &\!\!+\, 2\cdot |\, 2\, (1,1,1,1) \,|^2
\Cr  \end{array} \mbox{~~~~~~~~~~~~~~~~~~~~~~~}\ee
and
  \be  \begin{array}{ll} \mbox{{\normalsize $\zcec(E_{6,12}) =$ }}
 &  |\, (0,0,0,0,0,0)\, +\, (0,0,0,0,12,0) +\, (0,0,1,0,0,0)\,
  + (0,0,1,0,9,0) \Cr
 &  \quad +\, (0,0,2,0,3,0) +(0,1,0,0,5,2) +(0,1,0,2,1,0) +(0,2,0,0,1,0) \Cr
 &  \quad +\, (0,2,0,0,7,0) +(1,0,0,0,7,2) +(1,0,0,2,0,0) +(1,0,3,0,1,0) \Cr
 &  \quad +\, (1,1,1,0,3,1) +(1,1,1,1,1,0) +(1,2,0,0,5,1) +(1,2,0,1,0,0) \Cr
 &  \quad +\, (2,0,0,1,3,1) +(2,0,1,0,2,0) +(2,0,1,0,5,0) +(3,0,2,0,0,0) \Cr
 &  \quad +\, (3,0,2,0,3,0) +(3,0,1,1,1,1) +(3,1,0,0,2,1) +(3,1,0,1,3,0) \Cr
 &  \quad +\, (4,0,0,0,4,0) +(5,0,0,2,1,1) +(5,0,0,1,0,2) +(5,0,1,0,2,0) \Cr
 &  \quad +\, (7,0,0,2,0,0) +(7,0,0,0,1,2) +(9,0,1,0,0,0) +(12,0,0,0,0,0)\,|^2
\CR
 &\!\!\!+\, |\, (0,0,0,0,0,1)\, + (0,0,0,0,6,3)\, + (0,0,0,1,10,0)\,
  + (0,0,0,3,0,0) \Cr
 &  \quad +\, (0,0,4,0,0,0) +(0,1,0,0,8,1) +(0,1,0,1,0,0) +(0,1,2,0,2,0) \Cr
 &  \quad +\, (0,2,0,0,4,2) +(0,2,0,2,0,0) +(0,3,0,0,0,0) +(0,3,0,0,6,0) \Cr
 &  \quad +\, (1,0,1,0,4,1) +(1,0,1,1,2,0) +(1,1,0,0,6,1) +(1,1,0,1,1,0) \Cr
 &  \quad +\, (2,0,2,0,2,1) +(2,0,2,1,0,0) +(2,1,0,1,2,1) +(2,1,1,0,1,0) \Cr
 &  \quad +\, (2,1,1,0,4,0) +(3,0,0,0,3,1) +(3,0,0,1,4,0) +(4,0,0,2,0,2) \Cr
 &  \quad +\, (4,0,1,0,1,1) +(4,0,1,1,2,0) +(4,1,0,0,3,0) +(6,0,0,0,0,3) \Cr
 &  \quad +\, (6,0,0,1,1,1) +(6,0,0,3,0,0) +(8,0,0,1,0,1) +(10,1,0,0,0,0)\,|^2
\CR
 &\!\!\!+\, 2\cdot |\, 4\, (1,1,1,1,1,1) \,|^2 \Cr  \end{array}\ee

 \normalsize
\noindent
These results, which are extremely hard to even guess in any other way,
demonstrate the power of \qg\ symmetries quite convincingly.

\vskip 16mm \noindent
{\bf Acknowledgement.} It is a pleasure to thank T.\ Gannon for stimulating
discussions.

\version\versionno
\end{document}